\documentclass[twocolumn]{aastex631}
\usepackage{xspace}
\usepackage{multirow}
\usepackage{natbib}
\usepackage{mathtools}
\usepackage{comment}

\newcommand{\be}{\begin{equation}}
\newcommand{\ee}{\end{equation}}
\newcommand{\bea}{\begin{eqnarray}}
\newcommand{\eea}{\end{eqnarray}}

\newcommand{\jwst}{\textit{JWST}\xspace}

\defcitealias{harris_reinacampos2023} {Paper I}

\begin{document}

\title{\textit{JWST} Photometry of Globular Clusters in Abell 2744. II: luminosity and color distributions}

\author[0000-0001-8762-5772]{William E. Harris}
\affiliation{Department of Physics \& Astronomy, McMaster University, 1280 Main Street West, Hamilton, L8S 4M1, Canada}

\author[0000-0002-8556-4280]{Marta Reina-Campos}
\affiliation{Canadian Institute for Theoretical Astrophysics (CITA), University of Toronto, 60 St George St, Toronto, M5S 3H8, Canada}
\affiliation{Department of Physics \& Astronomy, McMaster University, 1280 Main Street West, Hamilton, L8S 4M1, Canada}

\correspondingauthor{WEH and MRC}
\email{harris@physics.mcmaster.ca and reinacampos@cita.utoronto.ca}



\begin{abstract}
Deep \jwst imaging of the giant galaxy cluster Abell 2744, at redshift $z=0.308$, is used to explore the features of its rich population of globular clusters (GCs), building on our initial survey of the system \citep{harris_reinacampos2023}.  We use the photometry of more than $10,000$ GCs over a three-magnitude range to construct the GC luminosity function (GCLF) and  color distribution (CDF). These results now specifically account for photometric incompleteness as a function of location relative to the five giant galaxies that dominate the gravitational potential of A2744.  The total GC population in A2744 is estimated at $N_{\rm GC} \geq 1.1 \times 10^5$, consistent with its high total mass.  We also directly compare the observed distributions with theoretical predictions for GC populations drawn from the recent EMP-\textit{Pathfinder} simulations \citep{reina-campos+2022b}, viewed at the same 3.5 Gyr lookback time as the cluster. 
The simulations match the observations well, with the only notable disagreement being that the simulations predict larger numbers of GCs at high metallicity formed after $z\simeq2$ than are seen in the data.
\end{abstract}



\section{Introduction}

Abell 2744 is an extraordinarily rich cluster of galaxies at redshift $z = 0.308$ and lookback time of 3.5 Gyr, with a total mass in the range $M_{\rm vir} \simeq 2 \times 10^{15} M_{\odot}$ \citep{boschin+2006,jauzac+2016,medezinski+2016}. It has been nicknamed Pandora's Cluster \citep{merten+2011} because of the depth and range of phenomena that it hosts, including a huge diversity of galaxies, a merging and assembly history seen in midstream, a complex distribution of strong and weak lensing of background sources, and an IntraCluster Medium (ICM) filled with X-ray gas and an old stellar population \citep[e.g.][]{owers+2011,jauzac+2016,medezinski+2016,kempner_david2004,boschin+2006,eckert+2015,jauzac+2016,merten+2011,kimmig+2023}.   It has already been the target for deep imaging and spectroscopic campaigns with the \textit{James Webb Space Telescope} (\jwst) with studies covering a wide range of goals \citep[e.g.][in addition to the works cited above]{bezanson22,weaver+2023,bergamini+2023,price+2023,atek+2023,ikeda+2023,furtak+2023}.

\begin{figure}
    \centering{
    \includegraphics[width=\columnwidth]{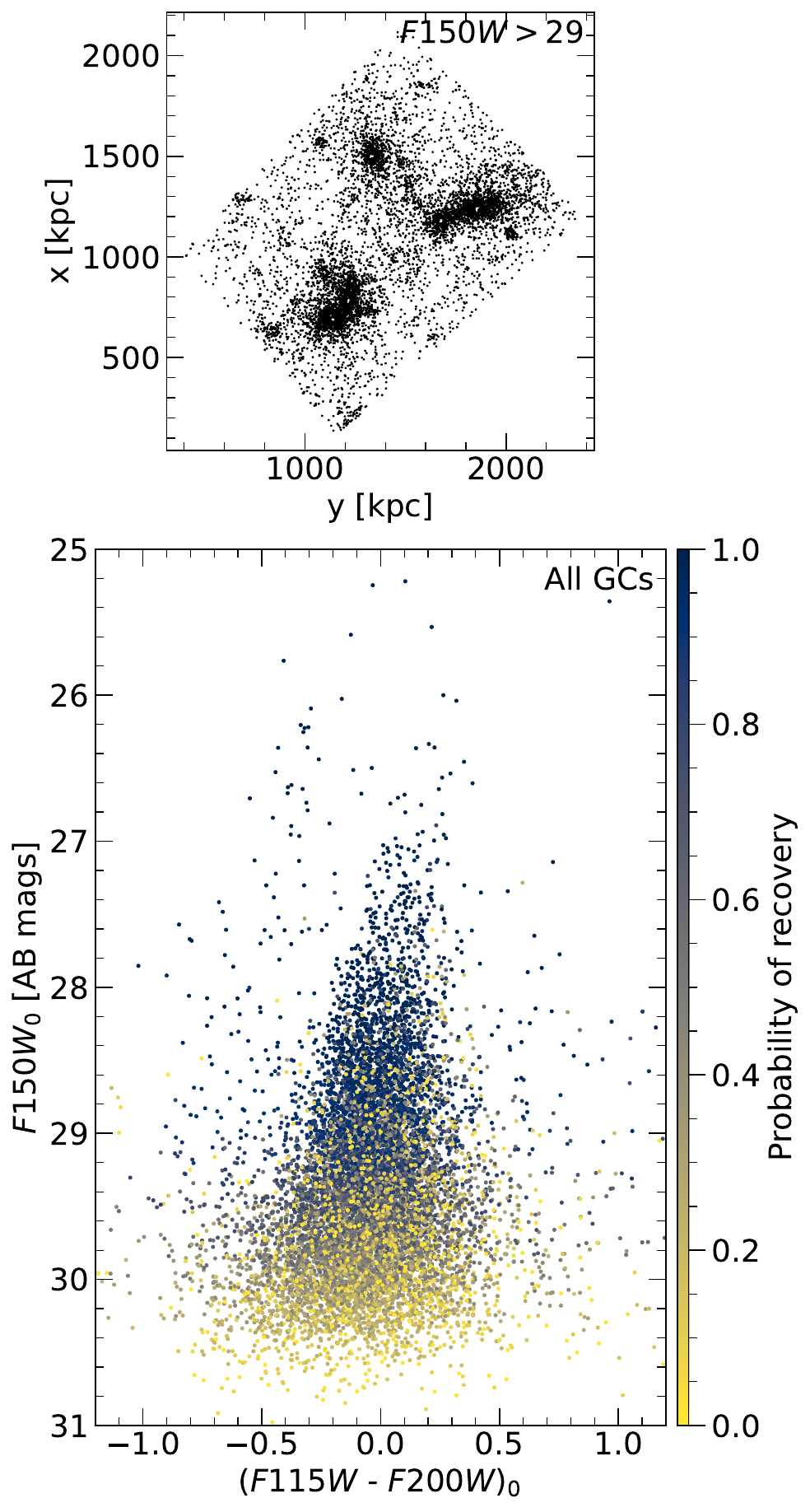}
  \caption{(\textit{Top}:) Locations of the $3,400$ GC candidates around the galaxies in the central Abell 2744 region brighter than $F150W = 29.0$, as measured on the \jwst NIRCam image mosaics. All objects plotted here are unresolved point sources, and their concentration around the major galaxies is evident. (\textit{Bottom}:) Color-magnitude diagram (CMD) for the entire GC sample in the A2744 field, showing apparent magnitude $F150W_0$ versus intrinsic color $(F115W-F200W)_0$ color-coded by recovery probability. Measurements are in the AB magnitude system, and include the cosmological K-corrections. At any given magnitude, there is a wide range of recovery probabilities because the GCs are drawn from a wide range of locations and thus local sky noise.}\label{fig:xy-cmd-plot} }
\end{figure}

In \citet[][hereafter \citetalias{harris_reinacampos2023}]{harris_reinacampos2023}, we presented photometry of more than 10,000 \emph{point sources} (unresolved, star-like objects) in the central region of A2744, using the UNCOVER mosaic images \citep{bezanson22} in the three \jwst NIRCam filters $F115W$, $F150W$, $F200W$.  The spatial distribution and CMD (color-magnitude diagram) are shown in Figure \ref{fig:xy-cmd-plot}. The point sources cluster strongly around the major A2744 galaxies and their satellites, and have a color range and lognormal magnitude distribution that clearly indicate we are looking deep into the globular cluster (GC) population around these galaxies, as well as in the ICM.  The brighter GCs may also overlap with a population of UCDs (Ultra-Compact Dwarfs), which would also be unresolved at the distance of the cluster (see \citetalias{harris_reinacampos2023}). This material opens up, for the first time, a deep look into the state of an entire large GC population at a significantly earlier stage of evolution, one-quarter of the way back to their origin.

The purpose of the present paper is to explore in more detail the CMD, the luminosity function (LF), and the color distribution function (CDF) of the GCs, going beyond the first-order assessment of \citetalias{harris_reinacampos2023}.  The outline of this study is as follows:  in Section~\ref{sec:lr}, artificial-star tests are described to assess the photometric recovery probability (i.e.~completeness) as a function of magnitude and location on the field. In Section~\ref{sec:gclf}, we analyze the completeness-corrected global LF for the GC population, along with its partner function the luminosity-weighted LF, with stronger tests of its match to a classic lognormal shape. In Section~\ref{sec:color-distribution}, we derive the completeness-corrected color distribution function (CDF). Section~\ref{sec:emp-pathfinder} introduces material from the EMP-\textit{Pathfinder} simulations of GC formation within galaxies, specifically selected to match the lookback time of A2744. Section~\ref{sec:comparisons} then compares the observed CMD, LF, and CDF with the predicted distributions from the simulated GC population, which show excellent first-order agreement. Section~\ref{sec:summary} summarizes the findings of this study.

As in \citetalias{harris_reinacampos2023}, the Planck 2015 cosmological parameters $H_0 = 67.8$ km s$^{-1}$ and $\Omega_{\Lambda} = 0.692$ \citep{planck2016} are adopted.  For the A2744 redshift of $z = 0.308$ the luminosity distance is $d_L = 1630$ Mpc or $(m-M)_0 = 41.06$. The foreground extinction (NED database) of $A_V = 0.036$ has a negligible effect on the NIR magnitudes and colors used here.  

\begin{figure}
\centering{
  \includegraphics[width=\hsize]{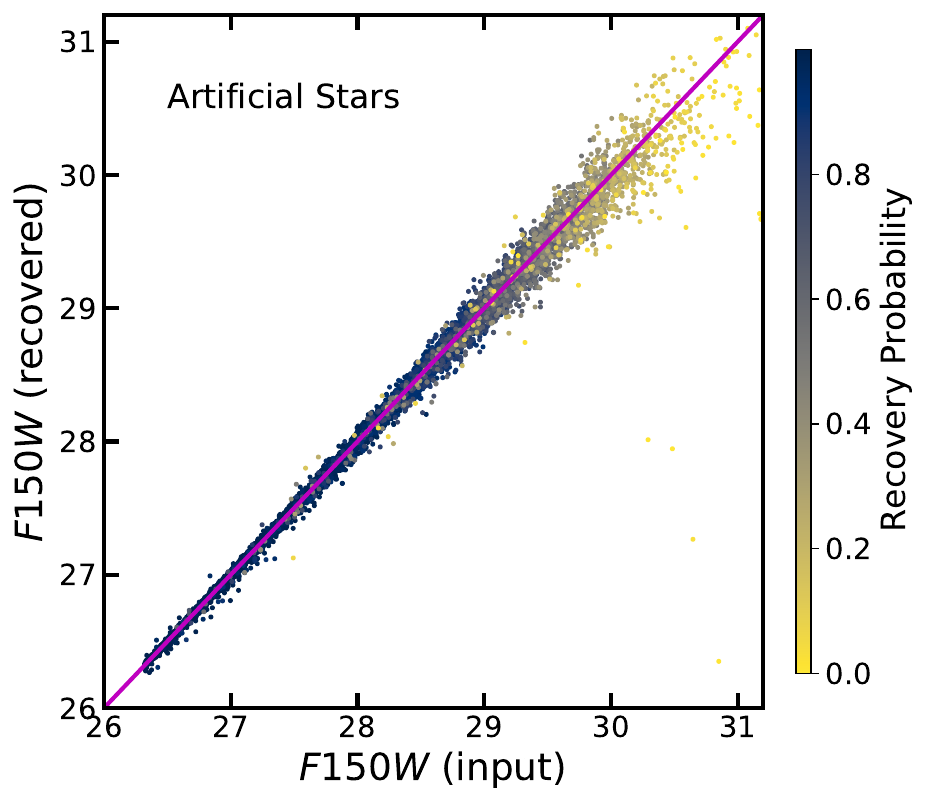}
  \includegraphics[width=0.93\hsize]{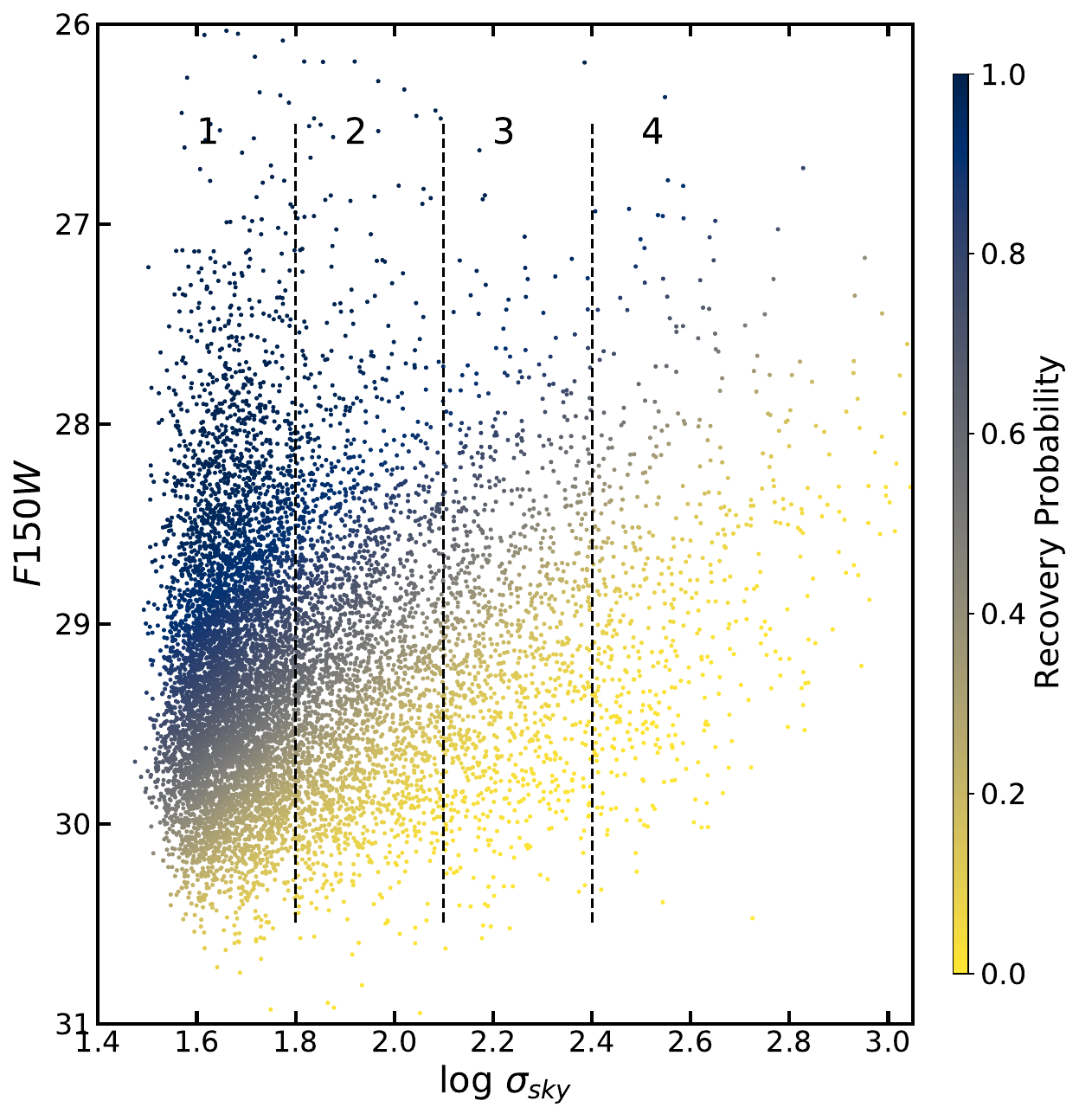}
  \caption{\emph{Upper panel:} Measured magnitude in $F150W$ of the artificial stars versus input magnitude and color-coded by their recovery probability. The 1:1 relation is shown as the solid line.  \emph{Lower panel:} Apparent magnitude in $F150W$ of the $10,617$ measured GC candidates as a function of local sky noise, $\log_{10} \sigma_{\rm sky}$ and color-coded by their recovery probability. The four zones of increasing sky noise as described in the text are labelled.}\label{fig:bias}
  }
\end{figure}

\section{Completeness Tests and Recovery Probability} \label{sec:lr}

The process of photometry and selection of the point sources in the A2744 field through the tools in \emph{daophot} \citep{stetson1987} is described in \citetalias{harris_reinacampos2023}. The measured magnitudes are on the ABMAG system as in \citet{weaver+2023}. After rigorous elimination of non-stellar objects, along with any sources that did not appear on all three of the $F115W$, $F150W$, $F200W$ images, the final list contained a total of $10,617$ point sources in the central area of A2744, the brightest $3,400$ of which are shown in Figure \ref{fig:xy-cmd-plot}. Most of the objects concentrate strongly around the major galaxies and their satellites, but several hundred are also found distributed widely in the ICM. The CMD for the entire set of measurements is shown in the bottom panel of Figure \ref{fig:xy-cmd-plot}, where the data include the cosmological $K$-corrections appropriate for GCs seen at redshift $z=0.3$ \citep[see][and \citetalias{harris_reinacampos2023}]{reina-campos_harris2024}.

To gauge the effective photometric limits of the data and the internal measurement uncertainties, artificial-star tests were performed additional to those in \citetalias{harris_reinacampos2023}. With the \emph{daophot/addstar} function, approximately $32,000$ fake stars covering the apparent magnitude range $26 - 34$ were inserted into the three images, in batches to avoid changes in the crowding levels anywhere. The frames containing the fake stars were remeasured with exactly the same procedure as for the real objects.  

If any inserted star is detected and measured in all three filters, \emph{and} it passes the culling tests for goodness of fit to the point spread function (PSF), it is said to be \emph{recovered}.  The \emph{completeness limit} is defined as the level at which the recovery probability $p$ equals 50\%. In these images, there are two main variables affecting the recovery completeness: magnitude (i.e.~faintness), and location relative to the major A2744 galaxies. Objects projected on areas of high surface brightness within the envelope of a big galaxy will be harder to detect, and the completeness limit will be correspondingly brighter. In \citetalias{harris_reinacampos2023}, the completeness was modelled only to first order as an average across the entire field and as a function of magnitude alone. Here, the recovery probability is modelled in more detail as an explicit function of both magnitude and location, and its effects on the GC luminosity function and color (metallicity) distribution are followed.

In several previous studies of GC systems around individual giant galaxies, projected galactocentric distance has been used as a proxy for surface brightness, and detection completeness $f(m)$ has then been derived in separate radial zones \citep[e.g.][]{whitmore+2010,faifer+2011,escudero+2015,escudero+2018,caso+2017,caso+2019,ennis+2020,dornan_harris2023}.  But in A2744 the cluster has three major subcenters dominated by five giant BCGs (Brightest Cluster Galaxies), and there are also several smaller but still significant satellite galaxies, several of which have their own GC populations.  The complexity of the total GC distribution makes it impossible to parametrize `location' in a simple way such as (e.g.) radial distance from any one of these galaxies.  However, the underlying variable affecting recovery probability is not radial location as such, but rather \emph{local sky noise}, which increases with surface brightness and crowding, and directly determines the detection threshold for the photometry.  The standard deviation of the local sky noise $\sigma_{\rm sky}$ is the `location' parameter that we opt to use here.

Following \citet{rosolowsky+2021} and \citet{harris_speagle2024}, we model the recovery probability $p$ with a logistic regression (LR) function, 
\begin{equation}
    p = \frac{1}{1+e^{-\mathbf{g}}} = \frac{1}{1+e^{-\mathbf{X}\ \mathbf{\beta}}}.
    \label{eq:logit}
\end{equation}
where $\mathbf{g}$ is the logit function, $\mathbf{X}$ is the vector of predictor variables, and the $\beta_i$ are the coefficients to be solved for
\citep[cf.][]{eadie+2022,harris_speagle2024}.

In terms of the neural network formalism of \citet{harris_speagle2024}, Eq.~(\ref{eq:logit}) is a single-layer, single-neuron model for the probability of recovery of any individual object with magnitude $m$ and local sky noise $\sigma_{\rm sky}$.  It can take this simple form because the data here do not show extra complications such as bilinear color-index dependence of the completeness limit, or (see below) significant nonlinear trends between input and measured magnitudes. 
The major advantage of treating the parameters within a single model of this form is that a recovery probability $p(m,\sigma_{\rm sky})$ can be assigned to every individual object in the list of measurements.

Specifically, the logit function adopted here is 
\begin{equation}
    g = \beta_0 + \beta_1 \cdot F150W + \beta_2 \cdot {\rm log}_{10} \sigma_{\rm sky} \, .
    \label{eq:lr}
\end{equation}
For the second parameter $\sigma_{\rm sky}$ we use the standard deviation of the sky pixel intensities within an annulus of 10 to 20 pixels ($\simeq 4 - 8$ fwhm of the PSF) around each object, as measured with the \emph{daophot/phot} aperture photometry function.  The numerical scale for $\sigma_{\rm sky}$ is arbitrary (only the ratios in different locations are important) but for convenience the data units per pixel on the original mosaic images were normalized to 10000 sec exposure. 

For the complete set of artificial stars, which cover the full desired range of magnitudes and sky noise, a maximum-likelihood solution was carried out in Python with \emph{statsmodels/scikitlearn}. The result for the coefficients is $\beta_1 = -2.59 \pm 0.04$, $\beta_2 = -5.37 \pm 0.15$.\footnote{The constant term $\beta_0$ has no physical meaning by itself since it depends on the adopted normalization of the intensity levels.}  The classification accuracy achieved by the solution is a satisfactorily high ${\rm CA}= 93.5~$per cent. This LR solution is then applied to the list of real objects to derive the recovery probability $p(m,\sigma_{\rm sky})$ for each one.

A technical issue to note is that for the artificial stars, the magnitude used in the LR solution is the \emph{input} magnitude, whereas for the real GCs, the \emph{measured} magnitude must be used as it is the only quantity available. However, there is no significant systematic bias between them, as shown in Figure \ref{fig:bias}a. Particularly for the critically important range $p > 0.5$ where the data have the highest reliability, the input and output magnitudes agree closely, and the 1:1 relation continues faintward even further.  

The increase in the internal measurement uncertainty towards fainter magnitudes is taken as the rms scatter of the points around the mean line.  The result is accurately described by a simple exponential curve, $\sigma_m = 0.01 + 0.07 e^{(m-29)}$ at magnitude $m$. 

\section{The GC Luminosity Function}\label{sec:gclf}

The next stage of this analysis is to derive GC luminosity distributions and color (metallicity) distributions (Sect.~\ref{sec:color-distribution}) that are fully corrected for completeness versus both magnitude and location and thus more systematically correct.

\subsection{Characterizing the GCLF}

The color-magnitude diagram of the real GC population, where the points are now color-coded by their individual recovery probabilities, is shown in Figure \ref{fig:xy-cmd-plot}b. It is evident that, at any given magnitude, a range of $p-$values is present, reflecting the effect of location and local sky level. 

This effect is shown more directly in Figure \ref{fig:bias}b, where $F150W$ magnitude is plotted directly against $\sigma_{\rm sky}$.  
Four zones in steps of factors of 2 in $\sigma_{\rm sky}$ are defined as marked out in Fig.~\ref{fig:bias}b, with boundaries at $\log_{10}\sigma_{\rm sky} = 1.8, 2.1,$ and $2.4$.  In Figure \ref{fig:cmd4}, the CMDs now divided into these four zones are shown, along with the spatial distributions of the same zones.  This graph demonstrates that higher sky noise is primarily associated with regions closer toward the centers of the major galaxies, but not exclusively so: other areas of increased sky noise are associated with satellite galaxies or even a few bright stars around the field.

To construct a completeness-corrected GCLF, for the present discussion we conservatively use only the data from Zone 1 ($\log_{10}\sigma_{\rm sky} \leq 1.8$). This region, which does cover most of the area of the field, clearly reaches the deepest magnitude for a given completeness value (Fig.~\ref{fig:bias}b). In principle, Zones 2--4 could be used to supplement the counts at the brighter magnitude levels. However, this method would necessarily assume that the  GCLF is independent of location in the potential wells of the five giant galaxies that dominate the GC counts. But the rate of dynamical evolution and mass loss of the GCs depends on radial location \citep[e.g.][]{gieles_baumgardt2008,lamers+2010,gnedin+2014,meng_gnedin2022}, so intrinsic differences in the LF should be expected between zones (though particularly at lower cluster masses). 

To generate the LF, each object is weighted by $(1/p)$ for its recovery probability $p$, and the resulting sums are shown in the left panel of Figure \ref{fig:lf}, binned in $0.2-$mag steps. The corrections to the raw count totals become quite significant for $F150W \gtrsim 29.$ The LF is then fitted with a Gaussian (lognormal) function, 
\begin{equation}
\phi(m) = \frac{N_{\rm tot}}{\sqrt{2 \pi} \sigma} \exp\left[\dfrac{-(m-m_0)^2}{2 \sigma^2}\right]
\label{eq:lf}
\end{equation}
where $\phi(m)$ is the (completeness-corrected) observed number of objects per unit magnitude. The total population over all luminosities $N_{\rm tot}$, the turnover magnitude $m_0$, and the standard deviation or LF dispersion $\sigma$ are the parameters to be solved for.  

As discussed in \citetalias{harris_reinacampos2023}, the observational limit is expected to fall short of the normal turnover point (peak frequency) of the GCLF. An unweighted nonlinear least-squares fit gives $m_0 = (31.86 \pm 0.28)$ mag for the turnover point, and $\sigma=(1.38 \pm 0.07)$ mag for the Gaussian dispersion. The quoted errors reflect only the internal fitting uncertainties, and are relatively small because the sample size is large and it covers a three-magnitude range. Adopting an unweighted fit deliberately avoids emphasizing either the high- or low-luminosity ends of the LF unduly. 

The solutions for $m_0$ and $\sigma$ are correlated if the data fall short of the true turnover, as is the case here \citep{hanes_whittaker1987,harris+2014}.  A better idea of the external uncertainties can be obtained by running a range of restricted 2-parameter solutions, assuming a value for $\sigma$ and solving only for ($N_{\rm tot}, m_0$).  These are listed in Table \ref{tab:gclf} for $\sigma = 1.1 - 1.5$ in 0.1-mag steps. The last line of the table gives the full three-parameter solution. As expected, the estimates for $m_0$ and $N_{\rm tot}$ are sensitive to the assumed dispersion $\sigma$. 
External constraints on $\sigma$ based on the observed GCLFs in other, nearby large galaxies \citep{villegas+2010,harris+2014} would favor values in the range $\sigma \simeq 1.2 - 1.4$, but this range would allow a $0.8-$magnitude range in the turnover point $m_0$ and  a factor of two in $N_{\rm tot}$.  

\subsection{Superluminous red objects:  GCs or UCDs?}

\begin{figure*}
    \centering{
	\includegraphics[width=\hsize]{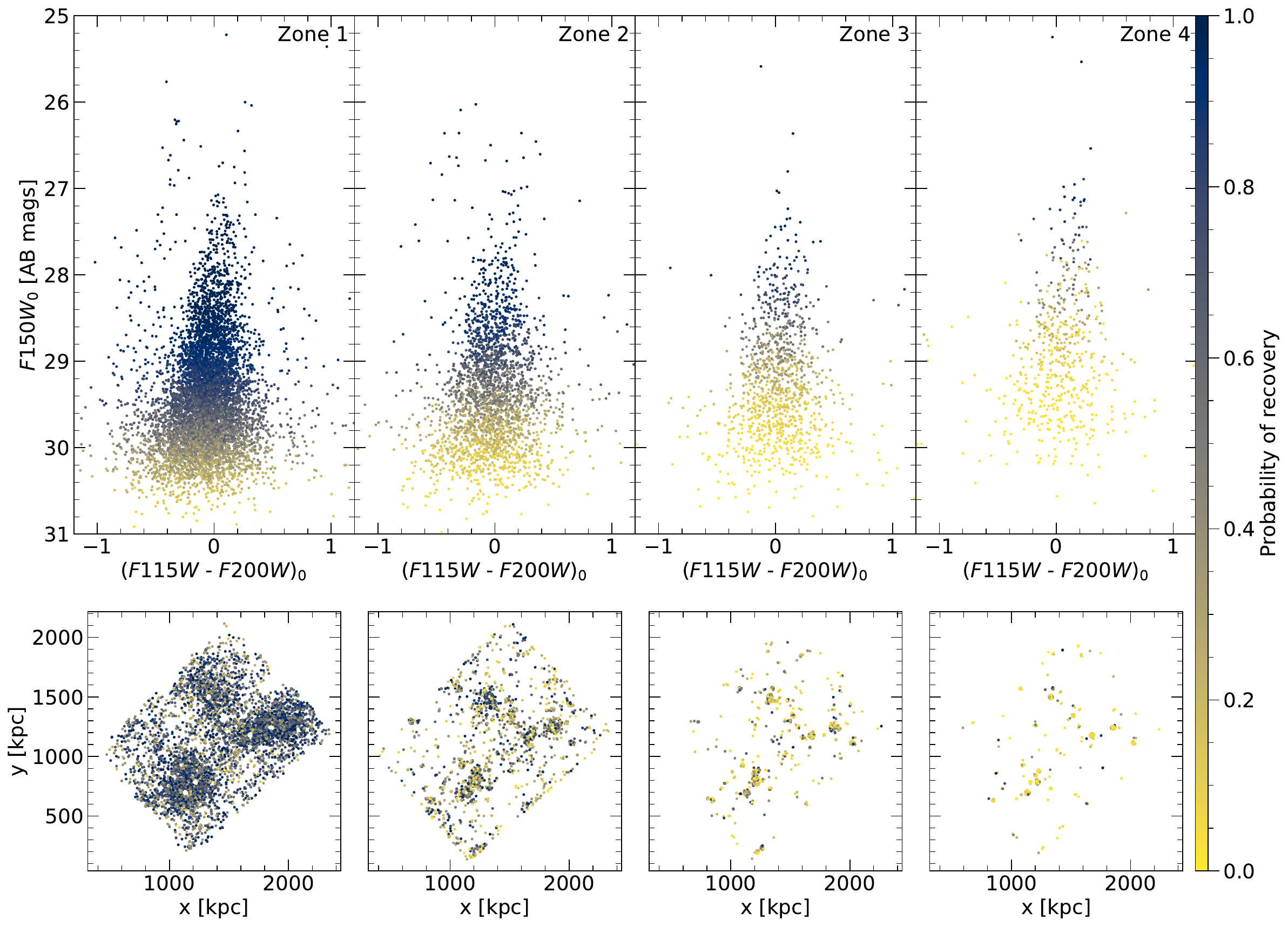}
    \caption{Properties of the GCs in Abell 2744 selected from increasingly noisier sky-noise zones (from left to right): CMDs for the GC candidates color-coded by recovery probability (\emph{upper row}), and their locations in the field (\emph{lower row}).}\label{fig:cmd4}
    }
\end{figure*}

In the CMDs of Fig.~\ref{fig:cmd4}, Zone 1 shows a striking upward extension of points for $M_{F150W} < -13.5$ above the redder half of the distribution, whereas in Zones 2 -- 4 there are relatively few of them.  Similar high-luminosity extensions of the red sequence appear in many (but not all) nearby BCG-type galaxies including NGC 1275, 3311, 4874, 6166, IC 4051, UGC 9799, UGC 10143, ESO383-G076, and ESO444-G046 \citep{mieske+2006,wehner_harris2007,madrid+2010,harris2023}.  In A2744, these luminous red objects are spatially more concentrated to the major BCGs than are the lower-luminosity GCs; the same appears to be true in other cases \citep{wehner_harris2007}.

\begin{figure}
    \centering{
    \includegraphics[width=0.90\hsize]{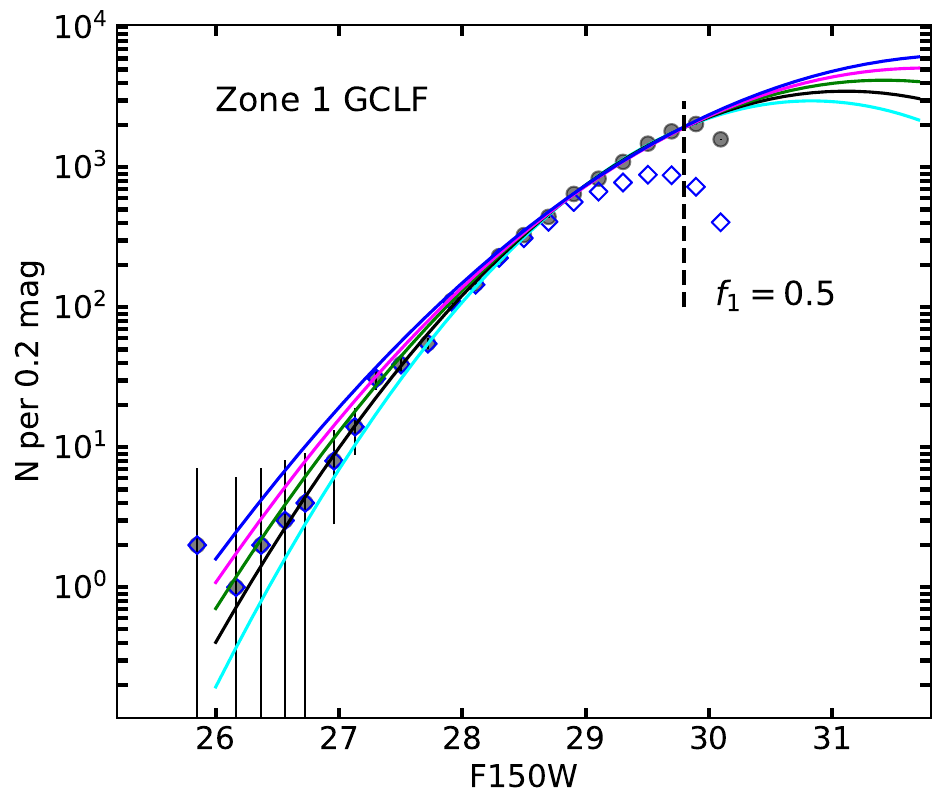}
    \caption{ GC luminosity function corrected for photometric incompleteness, as described in the text. The apparent magnitude $F150W$ here is not $K$-corrected. The completeness-corrected LF for Zone 1 ($\log_{10}\sigma_{\rm sky}\leq1.8$) is shown in the black filled circles, while the blue diamonds show the uncorrected raw counts. The approximate magnitude level at which the mean recovery fraction is $50~$per cent for Zone 1 is marked by the vertical dashed line at $F150W = 29.8$.  Errorbars are from $N^{1/2}$ count statistics.  The best-fit solutions for five different assumed values of the GCLF dispersion $\sigma=1.1$ (cyan), $1.2$ (black), $1.3$ (green), $1.4$ (red), and $1.5$ (blue) are shown by the solid lines. }\label{fig:lf}} 
\end{figure}

These `top end' GCs could have such high luminosities because they are \emph{more massive} than the bluer, more metal-poor GCs; or because they are systematically \emph{younger} (see Section \ref{sec:comparisons} below); or a combination of the two. This region of the CMD might also include a population of UCDs.  If so, this `extra' component does not show up in any obvious way as an excess at the bright end of the LF, since the basic lognormal LF shape is accurately followed all the way up through the very brightest objects (Fig.~\ref{fig:lf}).  Working only from the photometry and lacking any additional evidence about their intrinsic radii or masses, it is difficult to decide between the GC/UCD options (see the more extensive discussion in \citetalias{harris_reinacampos2023}).

\subsection{The total cluster population}

The LF fit described above gives the total GC population within Zone 1, but to estimate the true total population  the counts from all zones must be combined. The results from the LF indicate that Zone 1 contains very nearly half the total counts, at least for the brighter parts of the range.  Specifically, for $F150W < 28.50$, for which the completeness is higher than $50~$per cent for \emph{all} of Zones 1, 2, and 3, the completeness-corrected totals are $N = 979 \pm 32$ (Zone 1), $478 \pm 23$ (Zone 2), and $375 \pm 25$ (Zone 3).  Summing the three zones gives $N_{\rm tot} = (1.87 \pm 0.08) N_{\rm Zone1}$. With this ratio, and the \emph{assumption} that the GCLF shape is similar in the three zones, the $N_{\rm tot}$ results for the GCLF can be scaled to yield the total population over Zones $1$ -- $3$, which is $N_{\rm tot}(A2744) = 113,000 \pm 19,000$. As discussed in \citetalias{harris_reinacampos2023}, this total is very similar to the estimate of $N \simeq 100,000$ GCs for the comparably massive Coma cluster \citep{peng+2011}, but well below the initial estimate for  A2744  of $4 \times 10^5$ GCs by \citet{lee_jang2016} extrapolated from HST imaging that covered only the brightest $\simeq 1$ magnitude of the LF \citep[but see][for later downward revisions]{alamo-martinez+2017,harris+2017}.  
The total number of GCs as derived above, $N_{\rm tot}$, is in strict terms a lower limit, since it does not include the counts from the innermost Zone 4 (a small addition), or the remote outskirts of the cluster outside the boundary of the current field of view. As noted above, it also relies implicitly on the assumption that the GCLF shape is similar across all zones, which at some level should not be valid. The innermost zones should contain a more dynamically evolved GC population, though the most important effects will be on the numbers of low-mass clusters fainter than the turnover. Lastly, the estimated $N_{\rm tot}$ assumes that the GCLF continues to have a lognormal shape with the bright and faint halves having the same dispersion. Comparison with theoretical simulations (to be discussed in \S5 below) indicates that this assumption becomes increasingly less valid at larger lookback times, when the lower-mass GCs are still present in larger numbers than they are now \citep[cf. Fig.~5 in][]{reina-campos+2022b}. Pursuing this analysis further will require much more extensive modelling to build in the expected effects of dynamical evolution on the LF as a function of location around the major galaxies.

\begin{table}
\centering{
\caption{Best-fitting parameters to a lognormal describing the luminosity function of GCs in the Zone 1 ($\log_{10}\sigma_{\rm sky}\leq1.8$) noise level in Abell 2744.} \label{tab:gclf}
\begin{tabular}{cccc}
  \hline \hline
$\sigma$ & $N_{\rm tot}$ & $m_0$ ($F150W$) & $\chi_{\nu}$ \\
(AB mag) & (Zone 1) & (AB mag) & \\
   \hline
    $1.1$ & $41,140 \pm 1,132$ & $30.83 \pm 0.03$ & 1.203\\
    $1.2$ & $52,050 \pm 1,180$ & $31.11 \pm 0.03$ & 0.806\\
    $1.3$ & $68,320 \pm 3,120$ & $31.42 \pm 0.04$ & 1.058\\
    $1.4$ & $89,920 \pm 5,400$ & $31.76 \pm 0.05$ & 1.578 \\
    $1.5$ & $120,080 \pm 7,290$ & $32.12 \pm 0.07$ & 2.220 \\
    $1.39 \pm 0.07$ & $107,620 \pm 26820$ & $31.66 \pm 0.28$ & 0.946 \\ \hline  
\multicolumn{4}{l}{NB: The $m_0$ values are not K-corrected. They are } \\
\multicolumn{4}{l}{converted to absolute magnitude by $M_{F150W} = m_0 - 40.89$.}\\
\end{tabular}}
\end{table}

\begin{figure}
    \centering{
	\includegraphics[width=\hsize]{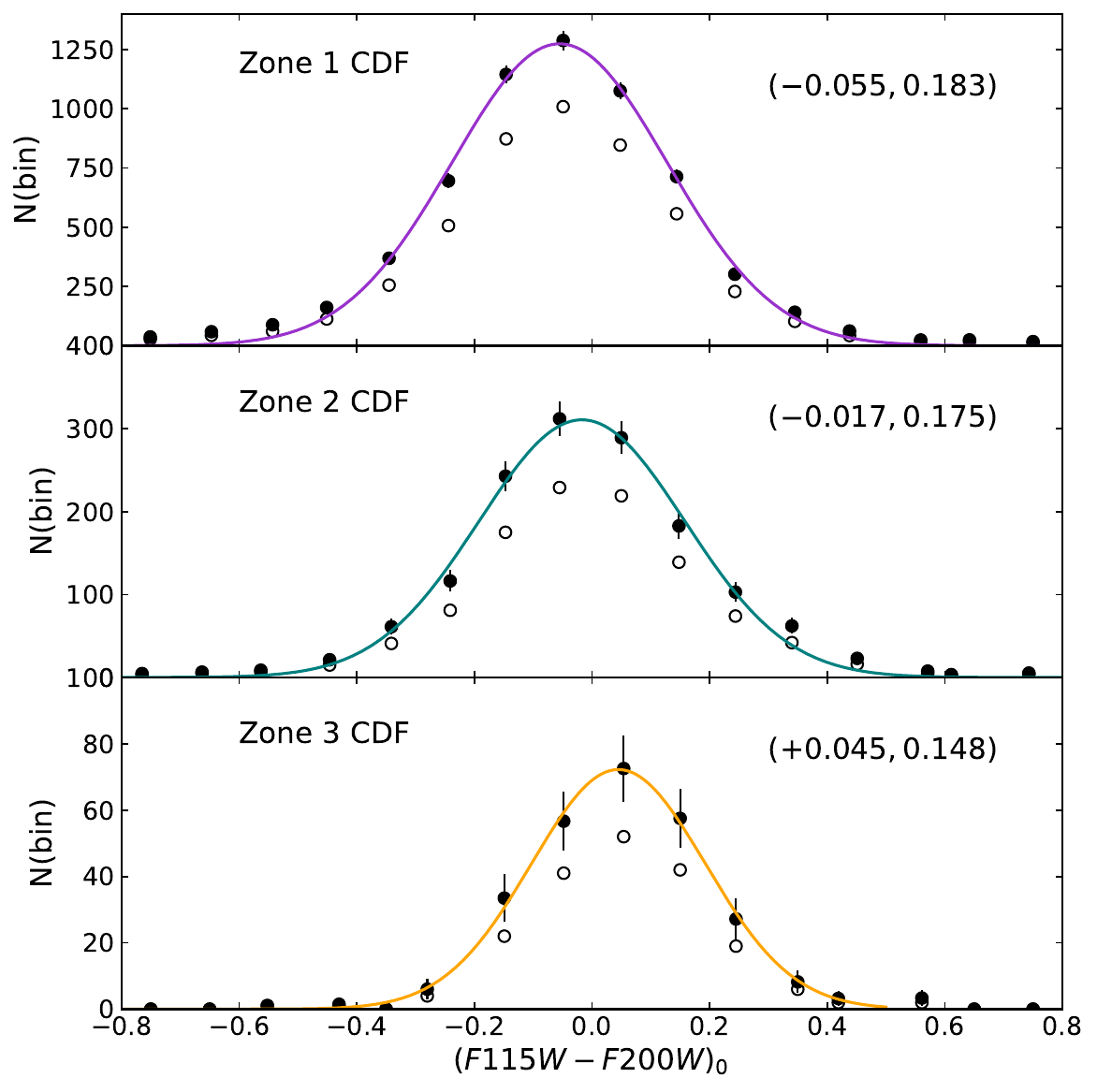}
    \caption{Color distribution functions (CDFs) for Zones $1$--$3$. In each panel the completeness-corrected totals per $0.1-$mag bin in color are plotted as the solid symbols with errorbars, while the raw counts are in the open symbols. The best-fit single Gaussian curve for each zone is overplotted with a solid line. The best-fit parameters ($c_0, \sigma_c$) are given at upper right in each panel.}\label{fig:cdf}}
\end{figure}

\section{Color and Metallicity Distributions}\label{sec:color-distribution}

We derive next the completeness-corrected color (metallicity) distributions of GCs. Figure \ref{fig:cdf} shows, for each of Zones $1$--$3$, the raw and completeness-corrected CDFs in ($F115W-F200W$). In each case the data included in the counts are restricted to the $F150W$ magnitude range where the completeness is higher than $50~$per cent. On average, the completeness-corrected values are $30~$per cent higher than the raw counts, but nearly independent of color.

The CDF in each zone is accurately matched by a single Gaussian function, with best-fit parameters as listed in the figure labels. Here, $c_0$ is the peak color index of the Gaussian and $\sigma_c$ the standard deviation in magnitudes. The simplest interpretation of the small but significant increase in peak color from Zones 1 through 3 is an increase in the mean metallicity of the GCs deeper in toward the centers of the major galaxies.  Converting the CDF parameters (see Section \ref{sec:emp-pathfinder} below for the conversion of metallicity to color) gives [m/H]$_0$ $= -0.64, -0.48, -0.21$ and dispersions $\sigma_{[m/H]} = 0.65, 0.63, 0.53$ for Zones 1--3.  
Metallicity gradients of this amount are conventionally observed in nearby large  galaxies \citep[cf.][for recent compilations and discussion]{liu+2011,forbes_remus2018,ko+2022,wu+2022,harris2023}, where GC heavy-element abundance is found to scale typically as $Z \sim R_{\rm gc}^{-0.3 \pm 0.1}$.  

From the extensive past literature (cf. the references cited above), we would expect to find that the CDF and its analog the MDF (metallicity distribution  function) should have an intrinsic bimodal-Gaussian form with identifiable blue (metal-poor) and red (metal-rich) subgroups.  Though a hint of that can be seen in the CMD of Fig.~\ref{fig:xy-cmd-plot} for the A2744 GCs, the relatively shallow  metallicity dependence of the color index ($F115W-F200W$), coupled with the increase in photometric measurement uncertainty with magnitude, have sufficiently broadened out the observed CDF to the point where any bimodal form has been obscured. We will analyze this point more quantitatively in Section \ref{sec:comparisons} below after introducing the simulated GC systems that will be compared with the data.

\begin{figure}
    \centering{
    \includegraphics[width=\hsize]{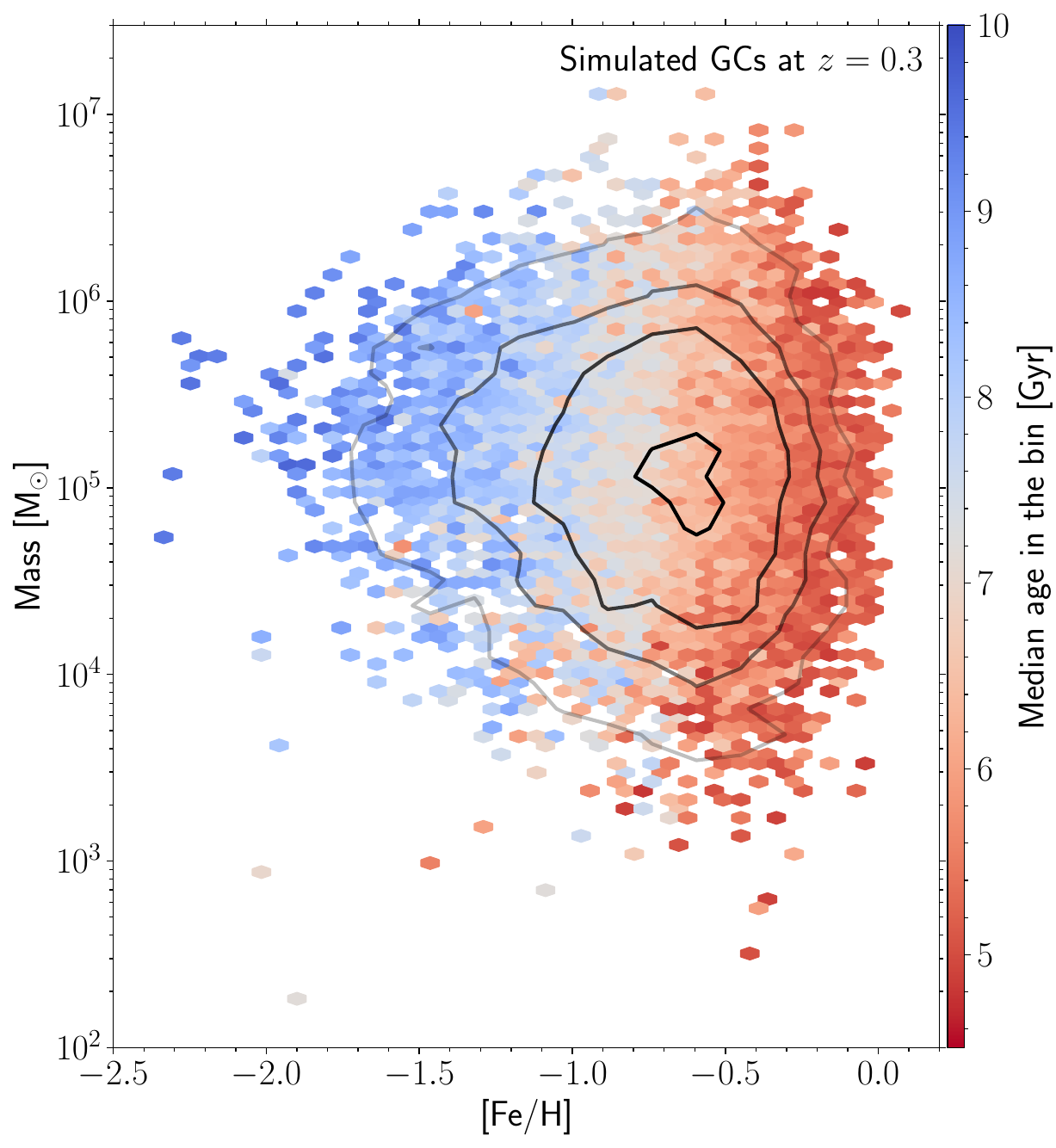}
    \caption{Mass as a function of metallicity for the simulated composite GC population at $z=0.3$ from the EMP-\textit{Pathfinder} simulations. The bins are color-coded by the median age of the GCs, and black solid lines of decreasing transparency enclose an increasing number of GCs.}\label{fig:simcmd}} 
\end{figure}

\section{The EMP-\textit{Pathfinder} Simulations}\label{sec:emp-pathfinder}

A new and more advanced stage of comparisons with theory is now becoming possible. Simulations of the formation and dynamical evolution of GCs within the hierarchical growth of galaxies are rapidly developing in scope and resolution, with many new insights and predictions for the distributions of GC mass, age, and metallicity \citep[e.g.][for recent examples]{choksi_gnedin2019,kruijssen+2019,el-badry+2019,reina-campos+2019,reina-campos+2022b,pfeffer+2023,chen_gnedin2024}.  In most cases, these previous studies have emphasized comparisons between theory and observational data for galaxies in the Local Universe, i.e.~at $z=0$.  The new data from \jwst, for the first time, now allow extension of these comparisons to different, significantly earlier epochs when GC systems were less evolved.

In particular, we focus here on comparing against results from the EMP-\textit{Pathfinder} simulations \citep{reina-campos+2022b}. This simulation suite models the concurrent evolution of a sample of 21 Milky Way-mass galaxies and their star cluster populations over cosmic time. The star clusters are modelled using a sub-grid approach and their initial mass and metallicity, as well as their subsequent mass evolution, are set by the galactic environment in which they reside. A critical component of these simulations is the presence of the physics leading to the multi-phase structure of the interstellar medium, which is critical to reproduce the mass-loss rate from tidal shocks against molecular clouds that molds their initial mass distribution into a lognormal shape.

Using these simulations, the evolution of the GC distributions by mass and metallicity can be explicitly traced as a function of redshift. A more complete discussion of these results will be presented in an upcoming study, but for the present purposes, we concentrate on the overall GC population in the simulated systems at $z=0.3$ for comparison with A2744. We show their predicted distributions of mass as a function of metallicity at a redshift of $z=0.3$ in Figure \ref{fig:simcmd}. 

To reduce the effects of statistical scatter and individual assembly histories, we combine the GC populations from our entire sample of 21 cosmological zoom-in Milky Way-mass simulations, which leads to a single population of $15,000$ GCs with a wide range of masses. Although the simulated population is a composite sample built from lower mass galaxies than BCGs, this approximation is still useful because the A2744 data also represent a composite population over many galaxies. In addition, current theory indicates that giant galaxies assembled from mergers and accretions of Milky Way-sized and smaller systems (cf. the discussions cited above), and the observed properties of GCs in present-day galaxies show extensive similarities and consistent trends across a wide range of galaxy masses \citep[e.g.][]{larsen+2001,peng+2008,villegas+2010,georgiev+2010,caso+2019,harris2023}. Thus for a first approximation, the comparisons of basic features such as the GCLF, MDF, and distribution in the CMD should be instructive. We also restrict the sample to GCs formed earlier than $z=1$ \citep[see sect.~6.2 in][for a detailed discussion]{reina-campos+2022b}, and we note that dynamical mass loss due to tidal shocks and evaporation is self-consistently tracked for each GC in the simulation.

Notably, the range of simulated GC ages (or their formation times, Fig.~\ref{fig:simcmd}) at $z=0.3$ runs from $\sim$ 5 to 10 Gyr, similar to the observed $4-5$ Gyr age spread of GCs within the Milky Way \citep{forbes_bridges2010,dotter+2011,vandenberg+2013}.  Said differently, the \emph{relative} age differences between GCs become increasingly important at higher redshift, particularly affecting their spread in luminosities. As seen in Fig.~\ref{fig:simcmd}, the bulk of the cluster population falls in the range $10^4-10^6~{\rm M}_\odot$.  The lower left corner of the diagram gives the appearance of a lack of low-mass, metal-poor clusters relative to the metal-rich side, but this is partly an illusion. As will be shown in \S\ref{sec:comparisons}, the actual shape of the luminosity function is closely similar between the metal-rich and metal-poor subgroups, with the major difference being simply the much larger total numbers of the metal-richer clusters.

For a different reason, there are also fewer low-metallicity clusters at \emph{high} mass (upper left corner of the figure). Any such clusters should be able to survive to the present time, but very low-[Fe/H] clusters were primarily formed in low-mass halos at the earliest stages of hierarchical growth. These small halos held proportionally lower amounts of gas from which to make clusters. In contrast, higher-metallicity clusters formed later in more massive halos with larger reservoirs of enriched gas from which it was easier to make massive clusters \citep[cf. the discussions of][]{harris+2006,choksi+2018,usher+2018,kruijssen2019}.

\section{Comparison against simulated data: CMD, GCLF, and MDF}\label{sec:comparisons}

\subsection{Transforming simulated quantities into observables}

The simulated GC population at $z=0.3$, as shown in Fig.~\ref{fig:simcmd}, opens an entirely new opportunity to compare theoretical predictions with direct observations for an evolutionary stage much earlier than the present epoch.  In order to do proper comparisons, the simulated physical properties (mass, age, metallicity) must be transformed into luminosity and color index so they can be matched to the A2744 observations. However, an important issue is that GC luminosity and color may be readily observable, but for distant galaxies GC age is a `hidden variable' not easily determined. Fig.~\ref{fig:simcmd} demonstrates that in any given region of the theoretical mass/metallicity plane, a significant range of age is present. By extension, similar ranges can be assumed to be present in the observed CMD for a real galaxy. For best accuracy, the transformations must be able to account for any combination of age and metallicity.

To develop appropriate conversions, we employed the PARSEC CMD3.7 stellar models\footnote{Available at http://stev.oapd.inaf.it/cgi-bin/cmd\_3.7} \citep{bressan+2012,marigo+2013} to generate a grid of $30$ single stellar population (SSP) models covering metallicities [m/H] = ($-2.0, -1.5, -1.0, -0.5, +0.0$) and ages of (5, 6, 7, 9, 11, 13) Gyr. These models have scaled-Solar abundances and were run with a \citet{chabrier2001,chabrier2005} lognormal initial stellar mass function to match what was used in the EMP-\textit{Pathfinder} simulations. Over the age range of $5$--$10~$Gyr, and with the \jwst/NIRCam filters, the absolute magnitudes and color indices change slowly enough with age and metallicity to allow relatively simple interpolation equations to be used.  For the $(F115W-F200W)_{AB}$ color index, a two-parameter solution matching the SSP models is given by
\begin{equation}
\begin{aligned}
    (F115W-F200W) &= 0.01 + 0.10 \log_{10}(\tau) \\
    &+ 0.26{\rm [m/H]} 
    + 0.033{\rm [m/H]{^2}}
    \label{eq:mdf}
\end{aligned}
\end{equation}
where $\tau$ is the GC age in Gyr.  Here the approximate conversion [m/H] $\simeq$ [Fe/H] + 0.3 is used to represent GC and halo stars \citep{gilmore_wyse1998,venn+2004,amarsi+2019,larsen+2022}. 
Similarly, the predicted absolute magnitude in $F150W_{\rm AB}$ corresponding to any cluster mass $M_{\rm GC}$ is well approximated by
\begin{equation}
\begin{aligned}
    M_{F150W,\rm AB} &= 3.15 + 1.24 \log_{10}(\tau)  + 0.094 {\rm [m/H]} \\
    &+ 0.062 {\rm [m/H]}^2 - 2.5 \log_{10} (M_{\rm GC}/M_{\odot})
\end{aligned}
\label{eq:absmag}
\end{equation}
These relations reproduce the models with a scatter of $\pm0.04$ mag in both color and absolute magnitude.  

\subsection{Color-Magnitude Diagram}

\begin{figure*}
    \centering{
	\includegraphics[width=0.90\textwidth]{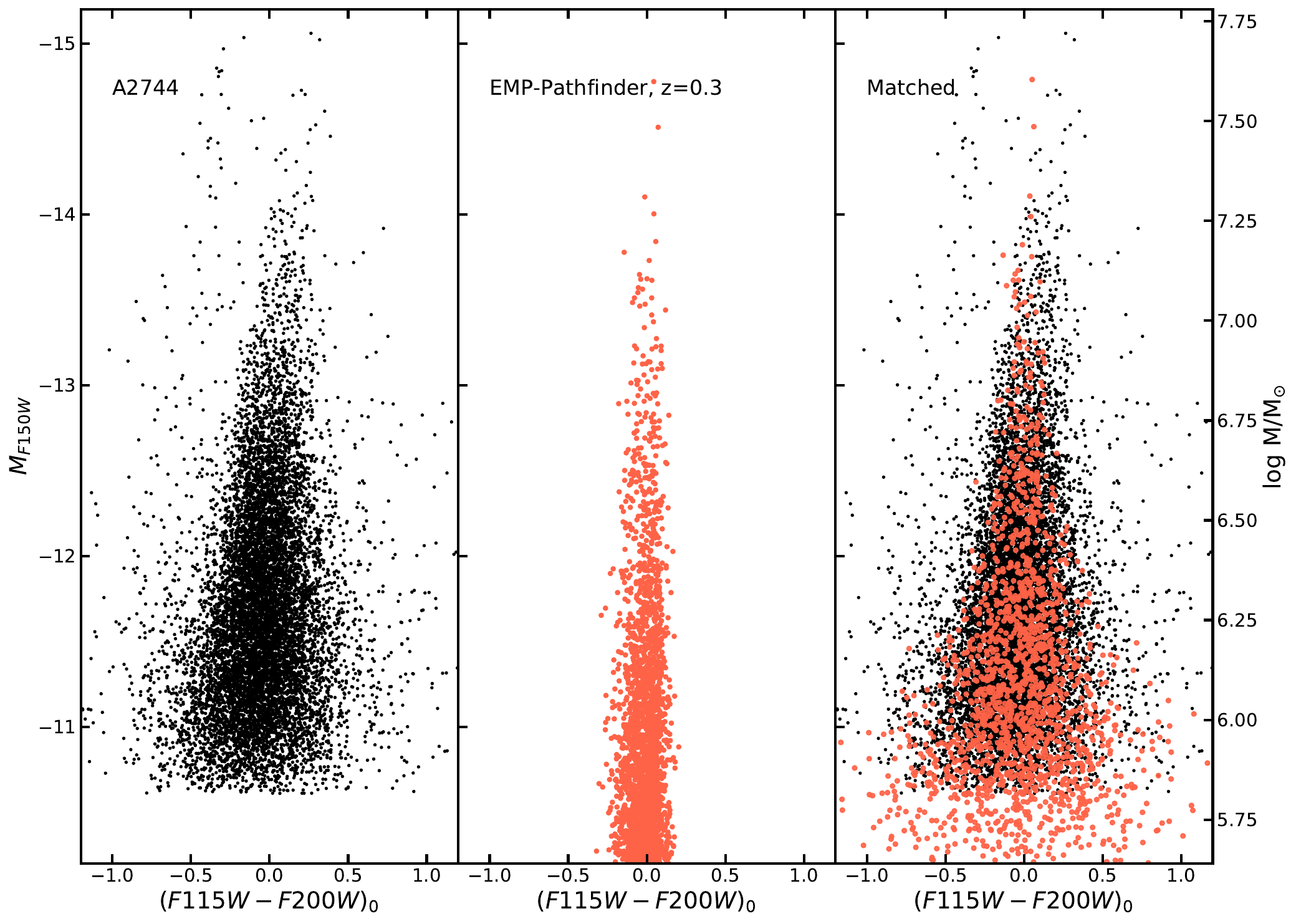}
    \caption{CMD in absolute magnitude vs color index for different samples: \emph{left panel:} CMD in absolute magnitude versus intrinsic color for the observed GCs in Abell 2744 present in all the sky noise zones, \emph{middle panel:} CMD for the composite GC population from the EMP-\textit{Pathfinder} simulations, converted from mass vs. metallicity to absolute magnitude vs. color index, and \emph{right panel:} observations (black points) matched with the transformed simulations (orange points). The simulated data have been broadened with the random uncertainties of measurement determined from the artificial-star tests, and truncated for masses $M \lesssim 10^{5.8}~{\rm M}_\odot$. Note that the mass scale shown at right is approximate, since conversion to luminosity depends in detail on both metallicity and age.}\label{fig:match_cmd} }
\end{figure*}

With the conversion relations in hand, the theoretical mass-metallicity diagram of Fig.~\ref{fig:simcmd} can be converted to the observational plane of $M_{F150W}$ vs.~($F115W-F200W)_0$. In successive panels, we show in Figure \ref{fig:match_cmd} the observed CMD in Abell 2744 combining all the sky noise zones, the simulated CMD from EMP-\textit{Pathfinder} transformed to the observational plane, and finally both sets of data combined. In the last panel, the theoretical CMD is broadened by the measurement scatter as determined from the artificial-star tests summarized in Section \ref{sec:lr}. Here, to roughly match the observations, the simulated GC distribution is truncated for $M \lesssim 10^{5.8}~{\rm M}_{\odot}$.  The overall match between the two distributions confirms that the increased spread in colors towards fainter magnitudes particularly for the range $M_{F150W} \gtrsim -12.5$ is well explained by the simple increase in measurement uncertainty. It should be noted that the simulations are drawn from a GC population an order of magnitude smaller than the real A2744 population, so the numbers of simulated objects in Fig.~\ref{fig:match_cmd} brighter than any given absolute magnitude are correspondingly lower.

\begin{figure*}
    \centering{
	\includegraphics[width=0.95\textwidth]{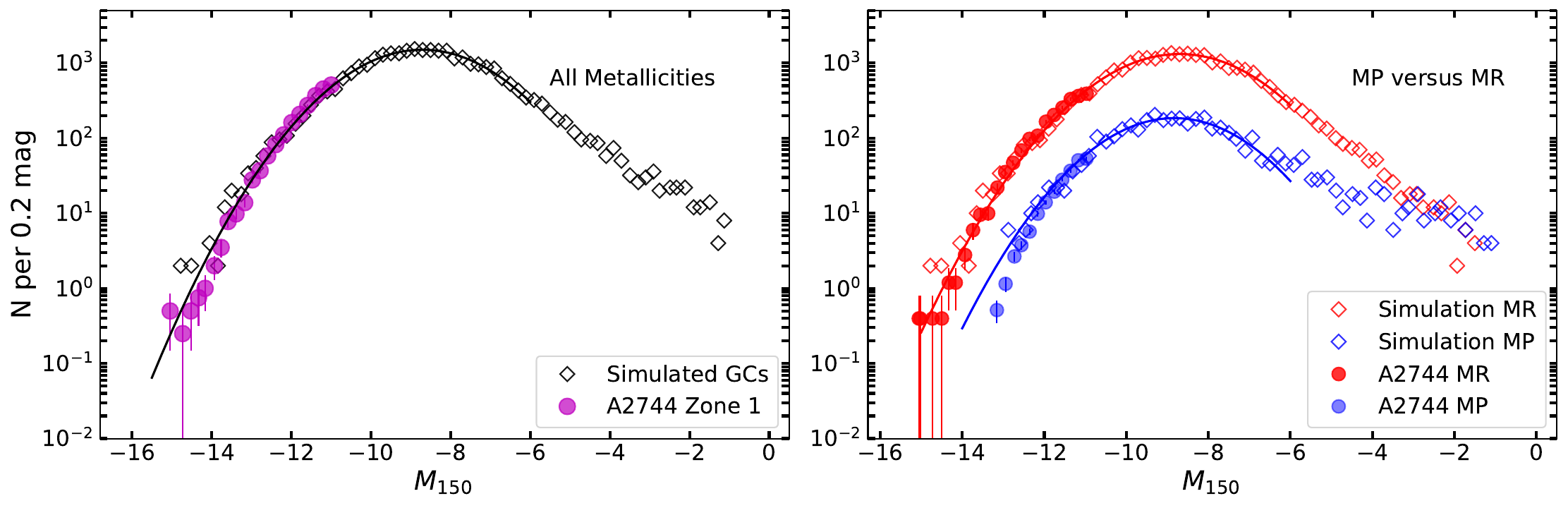}
    \caption{Comparison of the luminosity functions between the observed GC in Abell 2744 and the simulated population from the EMP-\textit{Pathfinder} simulations. The luminosity functions are shown as number of objects per $0.2$--mag bin versus absolute magnitude. \emph{Left panel:} Luminosity functions of the simulated GCs in EMP-Pathfinder (open diamond symbols), compared to the LF of the observed GCs within Zone 1 (filled dots with errorbars in magenta). The solid line is the best-fit Gaussian curve to the simulated data summarized in the text, and the data for A2744 is shifted vertically to fit the simulations. \emph{Right panel:} Luminosity functions of the metal-poor (blue markers) and metal-rich (red markers) GC populations. Open diamonds correspond to the simulated GC population, whereas filled circles show the observational data. Solid lines are the best-fit Gaussians to the simulated subsamples, and the data from the Zone 1 in A2744 is shifted to match the simulations.}\label{fig:lf_sim}} 
\end{figure*} 

\subsection{GC luminosity function}

The luminosity function can be addressed next. In Figure \ref{fig:lf_sim}a, the complete GCLF is shown for the composite population of simulated GCs at $z=0.3$. Here the masses of the simulated clusters have been converted to absolute magnitude $M_{F150W}$ with Eq.~\ref{eq:absmag}. The GCLF for the observed A2744 sample in Zone 1 is superimposed as the solid symbols with errorbars, arbitrarily shifted on the y-axis to test the similarity of LF shape. In general, the LF for the observed data defines a slightly steeper curve (i.e.~there are relatively too many simulated GCs at high luminosities, $M_{F150W} \lesssim -13$), but at fainter levels the two curves show close agreement.

The same analysis can be done splitting the data into two rough metallicity subgroups, as shown in Fig.~\ref{fig:lf_sim}b. The red symbols are for simulated GCs with [Fe/H] $> -0.8$ (i.e. primarily the yellowish bins in Fig.~\ref{fig:simcmd}a) and the blue symbols for [Fe/H] $< -0.8$. The corresponding dividing line for the A2744 observations is to some level ambiguous, but is set at $(F115W-F200W)_0 = 0.0$ following Fig.~\ref{fig:cdf}. The GCLFs of the simulated GCs are asymmetric, with the faint end having a clearly shallower slope than the bright end. This feature captures the earlier dynamical evolutionary state of the GC system, where the lower-mass clusters are preferentially removed over time and the mass distribution evolves from an initial approximate power-law down to a nearly lognormal shape at $z=0$. However, the brighter part of the LF ($M_{F150W} \lesssim -6$) is found to match a lognormal function quite closely. For the whole sample, the estimated turnover point at $M_{F150W} = -8.72 \pm 0.18$ corresponds to $F150W = 32.17$. Assuming a mean age of $8~$Gyr and mean metallicity [m/H] $\simeq -1$, that corresponds to a turnover mass $M \simeq 1.5 \times 10^5 M_{\odot}$, which is quite close to the observed value of $1-2 \times 10^5 M_{\odot}$ (depending on the assumed mass-to-light ratio) for nearby giant galaxies \citep[e.g.][]{jordan+2007,villegas+2010,harris+2014}.  Interestingly, the low-metallicity subsample of the simulations yields a turnover point of $M_{F150W} = -8.83 \pm 0.04$ that is $(0.13 \pm 0.05)$ mag brighter than the high-metallicity turnover, corresponding to $13~$per cent higher mass. Though barely significant, the difference is consistent with the $2-4$ older mean age of the metal-poor population, which has therefore had more time to experience dynamical removal of low-mass clusters.

For both the metal-rich and -poor subsamples, the simulated LF is noticeably broader at $\sigma \simeq 1.5$ than the $\sim 1.25-$mag value for the A2744 sample. For the luminosity range $M_{F150W} \gtrsim -13$, however, the shape of the simulated GCLF matches the observations well and a true turnover point near $M_{F150W} \sim -9$ cannot be ruled out. A final point of interest to be drawn from Fig.~\ref{fig:lf_sim}b is that the MR sample reaches higher luminosity than the MP ones do, because they are both younger and more massive than their MP counterparts.

\subsection{Color and metallicity distribution function}

Because the color index $(F115W-F200W)$ is a relatively shallow function of metallicity (Eq.~\ref{eq:mdf}) compared with more well known optical color indices \citep{peng+2006,harris2023}, the CDF is a much compressed version of the MDF (metallicity distribution function), as seen in Fig.~\ref{fig:match_cmd}.  The CDF in $(F115W-F200W)$ does resolve the intrinsic width of the MDF (see \citetalias{harris_reinacampos2023}), but as mentioned above, the  photometric measurement uncertainties are large enough to smooth over any fine structure such as the expected bimodal metallicity distribution.  Thus rather than attempting to transform the CDF into the MDF by inverting Eqn. \ref{eq:mdf}, instead we convert the metallicities from the simulated GCs to color indices while convolving them with the photometric measurement uncertainties as a function of magnitude.

\begin{figure*}
    \centering{
	\includegraphics[width=0.42\textwidth]{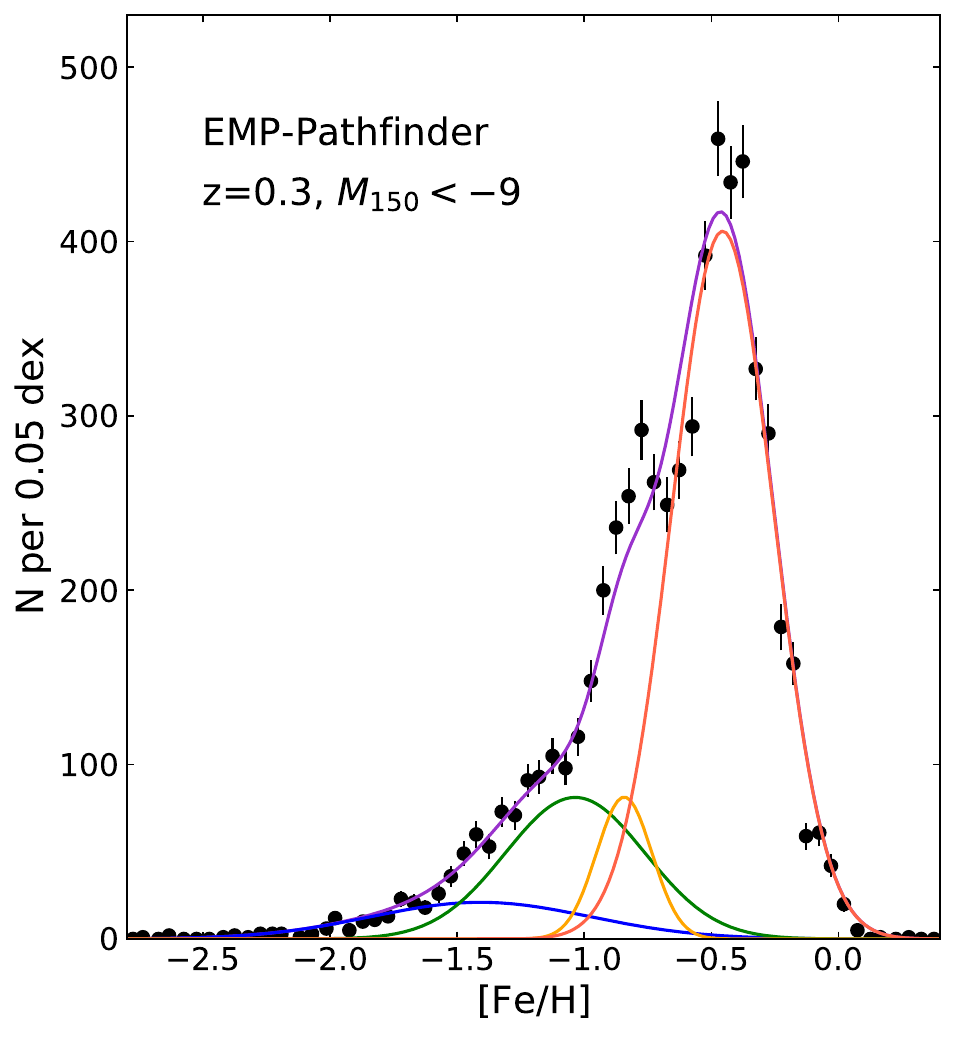}
 	\includegraphics[width=0.48\textwidth]{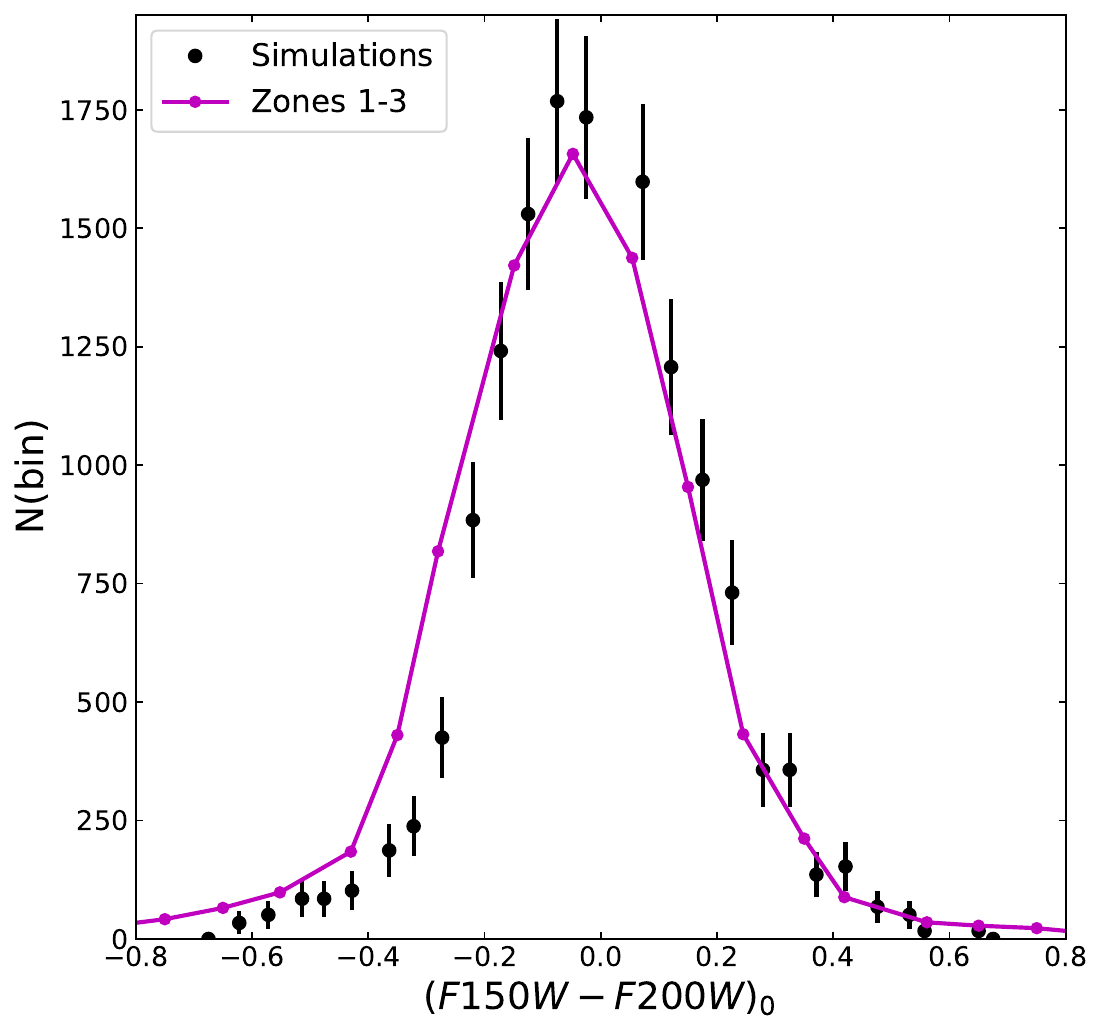}
    \caption{\emph{Left:} Metallicity distribution function (MDF) for the simulated GC population from the EMP-\textit{Pathfinder} simulations at $z=0.3$. The solid lines show a 4-component Gaussian fit to the distribution.  \emph{Right:} CDF for the combined observations from Zones 1--3 (magenta line with dots), compared with the predicted CDF from the EMP-\textit{Pathfinder} simulations (black points with errorbars). We include only clusters brighter than $M_{F150W} = -10.7$, which corresponds to the limit of the observed data. The simulated MDF from Fig.~\ref{fig:mdf_sim} has been transformed to the color index $(F115W-F200W)_0$ and then broadened by the observational measurement uncertainties and arbitrarily scaled vertically to yield the CDF.}\label{fig:mdf_sim}} 
\end{figure*}

The intrinsic MDF of the simulated population in its original form of [Fe/H] is shown in Figure \ref{fig:mdf_sim}. To generate the predicted CDF, we include only GCs more massive than $M \gtrsim 2 \times 10^5~{\rm M}_{\odot}$ ($M_{F150W} < -9$), which roughly correspond to the range above the LF turnover point.
The results of the transformed MDF are shown in Figure~\ref{fig:mdf_sim}.
Additionally, the broadened simulated distribution has been scaled vertically to roughly match the amplitude of the observed distribution of GCs in the sky noise zones $1$--$3$.

In a strict sense, the simulated MDF is unimodal (i.e.~there is only one peak, at [Fe/H] $\simeq -0.5$) but it is also quite complex, reflecting the assembly histories of the individual galaxies that were combined to make the total MDF.  Attempts to fit the MDF with the usual approach of Gaussian components quickly show that its shape is not well matched with single, double, or even triple Gaussians. Fig.~\ref{fig:mdf_sim} shows a sample fit with a 4-component Gaussian combination, where the four different modes have mean points at [Fe/H] $= (-1.41,-1.04, -0.84, -0.45)$ dex; standard deviations $\sigma_{\rm [Fe/H]} = (0.43, 0.28, 0.11, 0.20)$ dex; and fractions of the total $f = (0.075, 0.184, 0.073, 0.668)$.  
Adding more components to the fit can produce ever-closer matches to the data, but this would only be a numerical exercise. 

It appears that the classic bimodal form of the MDF that characterizes the GC systems in most nearby, zero-redshift galaxies \citep[e.g.][]{larsen+2001,peng+2006,harris2023} has its genesis between $z=1$--$2$, the range of formation times for most of the metal-rich clusters. At the same time, the simulations have relatively too many metal-rich GCs -- those with [Fe/H] $\gtrsim -1$. The metal-rich fraction $f_{\rm MR} = N(> -1)/N_{\rm tot}$ here is $75~$per cent\footnote{The constant rate of cluster formation from $z\simeq2$ in the simulations, which leads to an overproduction of metal-rich clusters, is the result of a numerical choice in the star formation scheme (see figs.~10--11 and the associated discussion in \citealt{reina-campos+2022b} and appendix~E in \citealt{pfeffer2018}).}, whereas in real systems it rarely reaches above $55~$per cent at any galaxy mass \citep[cf. figure 29 in][]{harris2023}.  We note, however, that no attempt has been made to select the simulated population by position (galactocentric radius) within their host galaxies.  Since the relative numbers of metal-rich clusters increase inward, selective removal by radius could result in an improved match with the observations.  In future work this step will be investigated more closely.

In Fig.~\ref{fig:mdf_sim}b, the shapes and positions of the observed and simulated CDFs are seen to agree well to first order, but at a finer level the simulated CDF lies slightly redward of the A2744 data. However, the offset is $\simeq 0.05$ mag, which is well within the combined uncertainties in the K-corrections, the photometric zeropoint calibration, the aperture corrections to the PSF-fitted photometry, and the \emph{daophot/allstar} measurements themselves, all of which are at the $\pm0.03$-mag level (see \citetalias{harris_reinacampos2023}). In brief, the overall match of this unusually deep set of observations to the predicted luminosities and colors generated from these state-of-the-art simulations is encouragingly close.

\section{Summary and Conclusions}\label{sec:summary}

Following our initial photometry of the globular cluster populations in the Abell 2744 galaxy cluster (see \citetalias{harris_reinacampos2023}), in the present paper we continue the analysis of the color-magnitude diagram, the luminosity function, and the color (metallicity) distribution of the star clusters. A new feature of this work is a comparison against the predicted GC populations from the EMP-\textit{Pathfinder} simulations, selected to match the look-back time of A2744 at $z=0.3$.

A brief summary of our findings is as follows:
\begin{enumerate}
    \item[(1)] Extensive artificial-star tests are used to model the recovery probability of the target objects as a simultaneous function of both magnitude and local sky brightness (sky noise). The results lead to the definition of distinct zones of sky brightness in which the color-magnitude diagrams and luminosity functions of the GCs have clearly different limiting magnitudes.  In `Zone 1', where the measurements reach the deepest and which occupies most of the A2744 field area, the 50\% completeness limit is at a magnitude $F150W = 29.8$.
    \item[(2)] The completeness-corrected luminosity function (GCLF) of the GCs in each of Zones 1--3 can be accurately matched by a standard lognormal function. Even in Zone 1, the data fall short of the predicted GCLF turnover (peak) point, so the estimated turnover point is sensitive to the assumed dispersion $\sigma$ of the LF. The results suggest $\sigma \simeq 1.25 \pm 0.1$ and a predicted turnover magnitude of $F150W \simeq 31.3 \pm 0.3$.   
    \item[(3)]  The total GC population over all zones and all magnitudes is $N_{\rm tot} \sim 113,000$, comparable with the estimates for the nearby Coma cluster. However, our estimate should be considered a strict lower limit given the fraction of the A2744 survey area covered and the uncertainty in the estimated GCLF turnover luminosity.
    \item[(4)] The recent EMP-\textit{Pathfinder} simulations of GC populations and their host galaxies at different evolutionary states (redshifts) have been used for direct comparison with the A2744 observed data. The very deep probe into the A2744 system has, for the first time, made it possible to compare predicted features like the CMD, the GCLF, and the MDF of these simulated populations with a real system at an age several Gyr in the past. 
    To first order the comparisons are encouraging. The theoretical mass vs. metallicity distribution for the simulated population accurately matches the observed A2744 CMD once the theoretical parameters are converted to magnitude and color, and suitably broadened by the measurement uncertainties. A key aspect of the transformations is that, at a lookback time of 3.5 Gyr, the \emph{relative} age differences between clusters are increasingly important, which means that the transformations to the observational plane must account for both metallicity and age.  The predicted GCLF turnover point is equivalent to a mass $M_{\rm GC} \simeq 1.2 \times 10^5~{\rm M}_{\odot}$, in close agreement with local GC systems.
    \item[(5)]  The intrinsic MDF of the simulated GC system is complex, but it has an incipient bimodal form. From the age distribution of the metal-richer clusters, these models suggest that the bimodality has its origins around redshift $z \sim 2$ and becomes more distinct as time goes on. One potential issue with the simulations is that relatively too many metal-rich clusters are present to adequately match the observations, by about a factor of two. 
\end{enumerate}

In future follow-up work, the A2744 GC population will be used to test recent work suggesting that the spatial distribution tracks the gravitational potential of the cluster as determined by the lensing map \citep[e.g.][]{alonso+2020,reina-campos+2022,reina-campos+2023,adamo+2023,lee22,diego23}.  

The EMP-\textit{Pathfinder} simulations and other contemporary theoretical models hold a rich set of results that will allow predictions for the expected distributions of GCs in systems at a wide range of evolutionary states. Deep imaging with \textit{JWST} has now brought many Gyr of lookback time within reach, opening a new era of mutual reinforcement between observation and theory.

\begin{acknowledgments}
As with \citetalias{harris_reinacampos2023}, we acknowledge the work of the UNCOVER team to produce the beautiful mosaic images used in this work. These images for Abell 2744 are publicly available at: \href{https://jwst-uncover.github.io/#}{https://jwst-uncover.github.io/\#}. 

The authors acknowledge the work done by Ben Keller, Diederik Kruijssen, Jindra Gensior, and Sebastian Trujillo-Gomez in the development of the EMP-\textit{Pathfinder} simulations. These simulations were run in the Graham supercomputing cluster from Compute Ontario, and in the BinAc cluster from the University of T\"ubingen. The research was enabled in part by support provided by Compute Ontario (\href{https://www.computeontario.ca}{https://www.computeontario.ca}) and Digital Research Alliance of Canada (\href{alliancecan.ca}{alliancecan.ca}). 

MRC gratefully acknowledges the Canadian Institute for Theoretical Astrophysics (CITA) Fellowship for support. This work was supported by the Natural Sciences and Engineering Research Council of Canada (NSERC).

\end{acknowledgments}

%

\vspace{5mm}
\facilities{\jwst (NIRCAM)}


\software{Daophot \citep{stetson1987},
          IRAF \citep{tody1986,tody1993}
          Jupyter Notebooks \citep{kluyver+2016}, 
          Matplotlib \citep{hunter2007},
          Numpy \citep{harris+2020b},
          PARSECv1.2 \citep{bressan+2012}
          }
          




\bibliography{mybib}{}
\bibliographystyle{aasjournal}



\end{document}